\documentclass[prl,aps,showpacs,twocolumn,preprintnumbers]{revtex4}
\usepackage{amsfonts}
\usepackage{amsmath}
\usepackage{amssymb}

\usepackage{dcolumn}

\usepackage{amsfonts}
\usepackage{amsmath}
\usepackage{amssymb}

\def\sn{\,{\rm sn}}
\def\dn{\,{\rm dn}}
\def\cn{\,{\rm cn}}
\def\sech{\,{\rm sech}}

\begin{document}

\title{ \textbf{Hidden nonlinear supersymmetry of finite-gap Lam\'e equation}}
\author{\textsf{Francisco Correa${}^1$, Luis-Miguel  Nieto${}^2$ and Mikhail S. Plyushchay${}^1$}\\
\vskip 0.1cm {\textit{${}^1$Departamento de F\'{\i}sica, Universidad
de Santiago de Chile, Casilla 307, Santiago 2, Chile
${}^2$Departamento de F\'{\i}sica Te\'orica, At\'omica y \'Optica,
Universidad de Valladolid, 47071, Valladolid, Spain \\\vskip 0.15cm
E-mails: francisco.correa@usach.cl, luismi@metodos.fam.cie.uva.es,
mplyushc@lauca.usach.cl}}}

\begin{abstract}
A bosonized nonlinear (polynomial) supersymmetry is revealed as a
hidden symmetry of the finite-gap Lam\'e equation. This gives a
natural explanation for peculiar properties of the  periodic quantum
system underlying diverse models and mechanisms in field theory,
nonlinear wave physics, cosmology and condensed matter physics.
\end{abstract}

 \maketitle


Supersymmetry \cite{SUSY}, as a fundamental symmetry providing a
natural mechanism for unification of gravity with electromagnetic,
strong and weak interactions, still waits for experimental
confirmation. On the other hand, in nuclear physics supersymmetry
was predicted theoretically \cite{IachelloNucl} and has been
confirmed experimentally \cite{NuclSUSY} as a \emph{dynamic}
symmetry linking properties of some bosonic and fermionic nuclei. It
would be interesting to look for some physical systems whose special
properties could be explained by a hidden  \emph{ordinary} (not a
dynamic)  supersymmetry.

In the present Letter we show that the quantum system described by
the finite-gap Lam\'e equation possesses a hidden supersymmetry. A
very unusual  nature of the revealed supersymmetry is that it
manifests as a \emph{nonlinear} symmetry of a \emph{bosonic} system
without fermion (spin) degrees of freedom. This means that we find
here a kind of a \emph{bosonized} supersymmetry giving a natural
explanation for peculiar properties of the periodic quantum problem
underlying many physical systems.

The Lam\'e equation first arose in solution of the Laplace equation
by separation of variables in ellipsoidal coordinates \cite{WW}, and
one of its early  applications was in the quantum  Euler top problem
\cite{KramItWang}. It plays nowadays a prominent role in physics
appearing in such diverse theories as crystals models in solid state
physics \cite{Suth,AGI}, exactly and quasi-exactly solvable quantum
systems \cite{Turb,FGR},  integrable systems and solitons
\cite{PerOl, Solitons}, supersymmetric quantum mechanics
\cite{SUSYper}, BPS monopoles \cite{BPS}, instantons and sphalerons
\cite{Dunne,Sphalerons}, classical Ginzburg-Landau field theory
\cite{MaiSte}, Josephson junctions \cite{Josj}, magnetostatic
problems \cite{DobRit}, inhomogeneous cosmologies \cite{RedSh},
Kaluza-Klein theories \cite{KK}, chaos \cite{chaos}, and preheating
after inflation modern theories \cite{PreHeat}. Most often, the
Lam\'e equation appears in physics literature in the Jacobian form
of a one-dimensional Schrodinger equation with a doubly periodic
potential,
\begin{equation}\label{Lame}
    H_j\Psi=E\Psi,\quad    H_j=-\frac{d^2}{dx^2}+j(j+1) k^2 \sn^2(x,
    k),
\end{equation}
where $\sn (x,k)\equiv \sn\, x$ is the Jacobi elliptic odd function
with modulus $k$ ($0<k<1$), and real and imaginary periods $4K$ and
$2iK'$,
 $K=K(k)$ is a complete elliptic integral of the first kind, and
 $K'=K(k')$, $k'^2=1-k^2$ \cite{WW,AbS}. A remarkable
property of this equation is that at integer values of the parameter
$j=n$, its energy spectrum has exactly  $n$ gaps, which separate the
$n+1$ allowed energy bands. The $2n+1$ eigenfunctions associated to
the boundaries $E_i(n)$, $i=0,1,\ldots, 2n$,
 of the allowed energy bands $[E_{0},E_{1}]$,
$[E_{2},E_{3}],\ldots, [E_{2n},\infty]$ are given by  polynomials
(`Lam\'e polynomials') of degree $n$ in the elliptic functions
$\sn\, x$, $\cn\, x$ and $\dn\, x$. These polynomials have real
periods $4K$ or $2K$, and the boundary energy levels $E_i(n)$ are
\emph{non-degenerated}. The states in the interior of allowed zones
are described by the quasi-periodic Bloch-Floquet wave functions
(which can be expressed in terms of theta functions \cite{WW}) of
quasi-momentum $\kappa(E)$,
$$
\Psi^\pm_E(x+2K)=\exp (\pm
i\kappa(E))\Psi^\pm_E(x).
$$
Every such interior energy level is
\emph{doubly degenerated}. For any non-integer value of the
parameter $j$, Eq. (\ref{Lame}) has an \emph{infinite} number of
allowed and prohibited zones.

The double degeneration of the energy levels is typical for a
quantum mechanical system with $N=2$ supersymmetry. But the presence
of $2n+1$ edge-bands singlet states in the $n$-gap Lam\'e equation
indicates on an unusual, nonlinear character \cite{AISP,PCM} of a
possible hidden supersymmetry. To reveal it, one notes that in the
limiting case $k=1$ we have
 $K=\infty$, $K'=\frac{\pi}{2}$,
and system (\ref{Lame})  reduces to the P\"oschl-Teller quantum
system given by the potential
$$
U(x)=-j(j+1)\sech^2 x+ j(j+1).
$$
The
latter, as it was shown recently in \cite{FP}, at $j=n$ possesses a
hidden polynomial supersymmetry \cite{AISP,PCM} of order $2n+1$
generated by the supercharges $Q_n$ and $\tilde{Q}_n=iRQ_n$,
\begin{eqnarray}\label{Supern}
    &[Q_n,H_n]=[\tilde{Q}_n,H_n]=0,
    \quad \{Q_n,\tilde{Q}_n\}=0,&\\
&Q_n^2=\tilde{Q}_n^2= P_{2n+1}(H_n),&\label{QPH}
\end{eqnarray}
where $P_{2n+1}(H_n)$ is some polynomial of the degree $2n+1$ of the
Hamiltonian $H_n$, $R$ is a reflection, $R\Psi(x)=\Psi(-x)$,
identified as the grading operator,
\begin{equation}\label{R}
    [R,H_n]=0,\quad \{R,Q_n\}=\{R,\tilde{Q}_n\}=0,\quad R^2=1,
\end{equation}
and $Q_n$ is a self-conjugate local differential operator of degree
$2n+1$. Based on this observation, first we note that in the trivial
case of the free particle system with $j=0$ characterized by the one
allowed (`conduction') band $[E_0(0),\infty]$, $E_0(0)=0$, the odd
first order differential operator $Q_0=-iD$, $D=\frac{d}{dx}$, is
identified as the supercharge. For the one-gap Lam\'e system
(\ref{Lame}) with $j=1$, let us look for the self-conjugate
 integral of motion $Q_1$,
$[Q_1,H_1]=0$,
in the form of the third order
differential operator.
A direct check shows that
\begin{equation}\label{P1}
    iQ_1=D^3+fD+\frac{1}{2}f',
\end{equation}
is the odd integral, $\{R,Q_1\}=0$.
Here
\begin{equation}\label{f}
    f:= 1+ k^2-3k^2\sn^2x,
\end{equation}
$f'=Df$. The double-periodic elliptic function $f$ with periods
$2K$ and $2iK'$ satisfies the elliptic curve equation
\begin{eqnarray}\label{f'}
    &(f')^2=\frac{4}{3}(a_1-f)(f-a_2)(f-a_3),&
\end{eqnarray}
whose characteristic roots are
\begin{eqnarray}
    &a_1=f(0)=1+k^2,\quad a_2=f(K)=1-2k^2,&\nonumber\\
    &a_3=f(K+iK')=k^2-2,& \label{roots}
\end{eqnarray}
$a_1+a_2+a_3=0$. Differentiation of Eq. (\ref{f'}) gives the
identities
\begin{equation}\label{fn}
    f''+2f^2=2b^2,\quad D^l(D^2+2f)f=0,
\end{equation}
where $b^2=-\frac{1}{3}(a_1a_2+a_1a_3+a_2a_3)= k^4-k^2+1$,
$l=1,2,\ldots$. Using these relations, one finds that
$Q^2_1=P_3(H_1)$,
$$
P_3(H_1)=(H_1-E_0(1))(H_1-E_1(1))(H_1-E_2(1)).
$$
The energies
\begin{equation}\label{Ej1}
E_0(1)=k^2,\quad E_1(1)=1,\quad E_2(1)=1+k^2
\end{equation}
correspond here to the eigenfunctions $\Psi^{(1)}_0=\dn\,x$,
$\Psi^{(1)}_1=\cn\,x$, $\Psi^{(1)}_2=\sn\,x$, which form a zero-mode
subspace (kernel) of the supercharge $Q_1$. The states in the
interior of the two allowed zones are described by the
quasi-periodic eigenfunctions
\begin{equation}\label{j1quasi}
    \Psi^\pm_E=\frac{H(x\pm\alpha)}{\Theta(x)}\exp(\mp xZ(\alpha)),
\end{equation}
where $H(x)$, $\Theta(x)$ and $Z(x)$ are the Jacobi Eta, Theta and
Zeta functions, while the parameter $\alpha$ is related to the
energy eigenvalue $E$ via the equation $E=\dn^2\alpha+k^2$, see Ref.
\cite{WW}. They are also the eigenstates of the  supercharge,
\begin{equation}\label{QB1}
    Q_1\Psi^\pm_E=\pm \sqrt{P_3(E)}\, \Psi^\pm_E.
\end{equation}

Assuming that the $j=2$ Lam\'e polynomials $\Psi^{(2)}_0=f+b$,
$\Psi^{(2)}_1=\cn\,x\dn\,x$, $\Psi^{(2)}_2=\sn\,x\dn\,x$,
$\Psi^{(2)}_3=\sn\,x\cn\,x$, $\Psi^{(2)}_4=f-b$ \cite{FGR} form a
zero-mode subspace of the fifth order integral $Q_2$, one finds
\begin{eqnarray}\label{Q2}
    &iQ_2=D^5+5fD^3+\frac{15}{2}f'D^2+\left(\frac{9}{2}f''+4f^2\right)D.&
\end{eqnarray}
In the same way for $j=3$ and $j=4$ a tedious calculation gives the
supercharges $Q_3$ and $Q_4$. We do not display their explicit form
here, but, instead, describe the general structure of the
supercharges corresponding to $j=0,1,2,3,4$. First, one notes that
if the derivative is assigned the homogeneity degree $d_h(D)=1$, in
accordance with Eq. (\ref{f'}) the function $f$ can be assigned
$d_h(f)=2$, and then $d_h(H_j)=2$ and $d_h(Q_j)=2j+1$. Every
supercharge has the leading term $D^{2j+1}$, the next term is of the
form $fD^{2j-1}$, and every subsequent term decreases the order of
the derivative on the right in one unit. The supercharges
corresponding to the even cases $j=0,2,4$ contain the last term to
be proportional to $D$. With this structure of the supercharges for
the first cases of $j=0,\ldots, 4$ at hands, we can fix now the form
of the supercharges in the generic case $j=n$. Let us present the
Hamiltonian operator in terms of the function $f$,
\begin{equation}\label{Hj}
    H_j=-D^2-h_j\left(f(x)-f(0)\right),\quad h_j=\frac{1}{3}j(j+1),
\end{equation}
and look for the supercharge in the form
\begin{eqnarray}
    &iQ_j=D^{2j+1}+\alpha_jfD^{2j-1}+\beta_jf'D^{2j-2}
    +(\gamma_jb^2+&\nonumber\\
    &+\delta_jf^2)D^{2j-3}+\lambda_jf'''D^{2j-4}+...,&
    \label{Qjgen}
\end{eqnarray}
where in coefficients associated to the factors $D^l$, $l\geq 0$, it
is necessary to include all the independent structures of
homogeneity degree $d_h=2j+1-l$ given in terms of $f$ and its
derivatives modulo identities (\ref{fn}). Requiring $[Q_j,H_j]=0$,
we can fix the first coefficients,
\begin{eqnarray}
    &\alpha_j=h_j\left(j+\frac{1}{2}\right),\,
    \beta_j=\alpha_j\left(j-\frac{1}{2}\right),\,
    \gamma_j=\frac{6}{5}\beta_j(j-1),&\nonumber\\
    &\delta_j=\frac{5}{36}\gamma_j(j-6),\quad
    \lambda_j=-\frac{5}{72}\gamma_j\left(j-\frac{3}{2}
    \right)(j-2).&
    \label{abg}
\end{eqnarray}
These coefficients allow us to find a recurrence relation for the
supercharges. We note that the kernel $K_{2n}$ of the supercharge
$Q_j$ with $j=2n$ is spanned by the $4n+1$ functions
\begin{equation}\label{Kerodd}
    \varphi_a\cdot (1,f,\ldots,f^{n-1}),\,\,\,a=1,\ldots,4;\,\quad f^n,
\end{equation}
$\varphi_1=\sn\, x\cn\, x$, $\varphi_2=\sn\, x\dn\,x$,
$\varphi_3=\cn\, x\dn\,x $, $\varphi_4=1$, which are linear
combinations of the Lam\'e polynomials of the degree $2n$. For
$j=2n+1$, the kernel $K_{2n+1}$ is formed
 by the $4n+3$ functions
\begin{equation}\label{Kereven}
     \chi_a\cdot (1,f,\ldots,f^{n-1}),\, a=1,...,4;\,\,\, \chi_a f^n,\, a=1,2,3,
\end{equation}
with $\chi_1=\dn\, x$, $\chi_2=\cn\, x$, $\chi_3=\sn\, x$,
$\chi_4=\sn\, x\cn\, x\dn\, x$, where for $n=0$ the states
proportional to $\chi_4$ are absent. In comparison with the kernel
$K_{j-2}$ of the supercharge $Q_{j-2}$, the kernel $K_j$ of the
supercharge $Q_j$ includes four additional states. These are
$\varphi_af^{n-1}$, $a=1,2,3$, and $f^n$ for $j=2n$, and
$\chi_af^{n-1}$, $a=1,2,3$, and $\chi_4f^{n-1}$ for $j=2n+1$.
Therefore, there should exist a relation
\begin{equation}\label{LamQ}
    Q_j=\Lambda_jQ_{j-2},
\end{equation}
where $\Lambda_j$ is a fourth order differential operator of
homogeneity degree $d_h(\Lambda)=4$ of the form (modulo the first
identity from (\ref{fn}))
\begin{equation}\label{Lamj}
    \Lambda_j=D^4+\tilde{\alpha}_jfD^2+\tilde{\beta}_jf'D
    +\tilde{\gamma}_jb^2+\tilde{\tau}_jf''.
\end{equation}
Using Eqs. (\ref{Qjgen}), (\ref{abg}), one finds the numerical
coefficients of the operator $\Lambda_j$
\begin{eqnarray}
    &\tilde{\alpha}_j=2j(j-1)+1\, ,\quad
    \tilde{\beta}_j=\frac{1}{3}j(4j^2-7)+\frac{3}{2}\, ,&\nonumber\\
    &\tilde{\gamma}_j=j^2(j-1)^2\, ,\quad \tilde{\tau}_j=\frac{1}{6}
    (j+3)(j+1)(j-1)^2.&
    \label{tildecoef}
\end{eqnarray}
The operator $\Lambda_j$ is not symmetric, and in the representation
(\ref{LamQ}) it serves to annihilate the additional zero modes of
$Q_j$ after application to them of the operator $Q_{j-2}$. There is
also an alternative recurrence representation of the supercharge,
$
    Q_j=Q_{j-2}\Lambda^\dagger_j.
$
The Hermitian conjugate operator $\Lambda^\dagger_j$ acts
invariantly on the kernel of $Q_{j-2}$,
 $\Lambda^\dagger_j: K_{j-2}\rightarrow K_{j-2}$,
 and transforms four additional states of the kernel of $Q_j$ into
 some linear combinations of the states of $K_{j-2}$.
Relation (\ref{LamQ}) (or,  $Q_j=Q_{j-2}\Lambda^\dagger_j$) allows
ones to calculate $Q_j$ for arbitrary even and odd values of $j$
proceeding from the explicitly displayed supercharges $Q_2$ (or,
$Q_0$) and $Q_1$.

With the fixed form of the integral $Q_j$, let us discuss the
general structure of a hidden polynomial supersymmetry. Since
$Q_j$ is a self-conjugate odd local differential operator, one can
introduce another, nonlocal supercharge, $\tilde{Q}_j=iRQ_j$.
$H_j$ and $R$ are commuting self-conjugate operators. Let
$\Psi_E^\pm$ be their common eigenstates,
$H_j\Psi_E^\pm=E\Psi_E^\pm$, $R\Psi_E^\pm=\pm\Psi_E^\pm$. Since
$Q_j$ commutes with $H_j$ and anticommutes with $R$, there exist
some linearly independent combinations $\Psi_{E,q}$ and
$R\Psi_{E,q}$ of $\Psi_E^+$ and $\Psi_E^-$ such that
\begin{equation}\label{Qeq}
    Q_j\Psi_{E,q}=q(E)\Psi_{E,q},\quad
    Q_jR\Psi_{E,q}=-q(E)R\Psi_{E,q}.
\end{equation}
Then, the states $\Psi_{E,q}$ and $R\Psi_{E,q}$ are the eigenstates
of $Q^2_j$ and $\tilde{Q}_j^2$ with the same eigenvalue $q^2(E)$,
and, hence, the same is valid for the states $\Psi^\pm_E$,
$Q^2_j\Psi^\pm_E=\tilde{Q}_j^2\Psi^\pm_E= q^2(E)\Psi^\pm_E$. All the
states $\Psi^\pm_E$ corresponding to the allowed zones constitute
the basis in the class of Bloch functions (including the periodic
and antiperiodic ones). Hence, we get the operator equality
$Q_j^2=\tilde{Q}_j^2=q^2(H_j)$. The operator $Q_j^2$ is a local
differential operator of degree $4j+2$. This means that the operator
$q^2(H_j)$ is a polynomial of degree $2j+1$ of its argument, i.e.
$q^2(H_j)=C(H_j-c_0)(H_j-c_1)\ldots(H_j-c_{2j})$, where $C$ is a
real constant, while $c_i$ are real, or some pairs of them could be
mutually conjugate complex numbers. Comparing the coefficients in
$Q_j^2$ and $q^2(H_j)$ before the operator $D^{4j+2}$, we find that
$C=1$. Then, applying the operator $q^2(H_j)$ to the $2j+1$
(anti)periodic eigenstates of the Hamiltonian $H_j$ corresponding to
the boundaries of allowed zones $E_i(n)$, and remembering that the
same states constitute the kernel $K_j$ of the supercharge $Q_j$, we
find that the set of the constants $c_i$  coincides with the set of
the boundary eigenvalues $E_i(n)$, $i=0,\ldots, 2n$. Thus, we have
shown that $Q_n^2=\tilde{Q}_n^2=P_{2n+1}(H_n)$, where
\begin{eqnarray}\label{QHPL}
     &{P}_{2n+1}(E):=\prod_{i=0}^{2n}(E-E_i(n))&
\end{eqnarray}
is the Lam\'e spectral polynomial. The nontrivial odd integrals
generate the order $2n+1$ \emph{polynomial superalgebra} being the
hidden symmetry of the \emph{bosonic} system (\ref{Lame}).

The nonlinear character of the local supercharges and supersymmetry
of the system (\ref{Lame}) is reminiscent to a nonlinear symmetry of
a particle in a Coulomb potential generated by the
Laplace-Runge-Lenz vector integral, and to that of an anisotropic
oscillator with commensurable frequencies \cite{Boer}. Let us
clarify the dynamical picture underlying the hidden nonlinear
supersymmetry having in mind the analogy with the anisotropic
oscillator. Consider the one-gap case. The Hamiltonian $H_1$ can be
factorized in three possible ways:
\begin{eqnarray}\label{H1fac}
    &H_1=A_{d}^\dagger A_{d}+k^2=A_c^\dagger A_c+1=A_s^\dagger
    A_s+1+k^2,&
\end{eqnarray}
where $A_d=D-(\ln \dn\, x)'$, $A_d\dn\, x=0$, and $A_c$ and $A_s$
have a similar structure in terms of the $\cn\, x$ and $\sn\, x$.
Write the Heisenberg equations of motion of
 $A_d$ and $A_s^\dagger$,
\begin{equation}\label{AdAs}
    i\dot{A}_d=\omega_d(x)A_d,\quad
    i\dot{A}_s^\dagger=-A_s^\dagger\,
    \omega_s(x),
\end{equation}
 $\omega_d(x)=-2(\ln \dn\, x)''$, $\omega_s(x)=-2(\ln
\sn\, x)''$. Define the operator $A_{s/d}=D-(\ln \sn\, x)'+(\ln\dn\,
x)'$ , for which
\begin{equation}\label{Asd}
    i\dot{A}_{s/d}=\omega_s(x)A_{s/d}-A_{s/d}\omega_d(x).
\end{equation}
Then the relation
\begin{equation}\label{Qfact}
    iA_s^\dagger A_{s/d}A_d=Q_1
\end{equation}
 gives us one of the six possible
factorizations of the supercharge (\ref{P1}). Note that operators
$A_c$, $A_s$, $A_{s/d}$ and associated instant frequencies have
singularities on a real line, which cancel in the  $H_1$ and $Q_1$.
For $j=n>1$ the same dynamical mechanism underlies the supercharge
structure and its possible factorizations.

We conclude that the physical systems associated with the $n$-gap
Lam\'e equation possess a hidden bosonized nonlinear supersymmetry.
It is behind the double degeneration of the energy levels in the
interior of the allowed bands and the singlet character of the
$2n+1$ edge-bands states. The latter form a zero-mode subspace of
the local supercharge $Q_j$ (as well as of the nonlocal one,
$\tilde{Q}_j$) being a differential operator of degree $2n+1$.
Taking into account the parity of the states (\ref{Kerodd}) and
(\ref{Kereven}), one finds that the system (\ref{Lame}) with any
$j=n$ is characterized by the  Witten index \cite{Witten} equal to
$1$. The information on the transfer matrix can also be extracted
from the structure of its hidden supersymmetry. The detailed
analysis of this aspect  will be presented in a separate
publication.

In the limit $k=1$, $\sn\,x=\tanh\, x$, $\cn\,x=\dn\,x=\sech\, x$,
the valence bands $[E_0,E_1],\ldots, [E_{2n-2},E_{2n-1}]$ shrink,
and two boundary states of a valence band transform into one bound
state of the related P\"oschl-Teller system. As a result, the
kernel of the supercharges of the latter system is constituted not
only by the bound eigenstates and the lowest eigenstate from the
continuous part of its spectrum, but also should include some $n$
unbounded states. The discussion  of this limit of the  Lam\'e
equation corresponding to the P\"oschl-Teller problem,  and its
relation to the bound state Aharonov-Bohm and the Dirac delta
potential systems from the viewpoint of the hidden supersymmetry
will be presented elsewhere.

A simple shift of the argument in the Lam\'e equation for a half
of the real period of the Hamiltonian $H_j$, $x\rightarrow x+ K$,
gives the isospectral doubly-periodic system
\begin{eqnarray}\label{Iso}
    \tilde{H}_j=-D^2+j(j+1)(1-k'{}^2\dn^{-2}(x,k)).
\end{eqnarray}
At $j=n$ it has a hidden polynomial supersymmetry generated by the
supercharges $Q_j$ and $\tilde{Q}_j$, whose explicit structure can
be obtained by applying the Jacobi functions identities
$\sn\,(x+K)=\cn\, x/\dn\, x$, $\cn\,(x+K)=-k'\sn\, x/\dn\, x$,
$\dn\, (x+K)=k'/\dn\, x$ to the supercharges of the system
(\ref{Lame}). It would be interesting to look for the hidden
polynomial supersymmetry in other finite-zone double periodic
quantum systems.

Finally, it would be interesting to clarify the role played by the
revealed hidden nonlinear supersymmetry of the finite-gap Lam\'e
equation in the periodic Korteweg-de Vries equation theory
\cite{Solitons} and in the periodic relativistic field theories
\cite{Dunne,Sphalerons,MaiSte} to which system (\ref{Lame}) is
intimately related.

\vskip 0.1cm
The work has been supported partially by the FONDECYT
Project 1050001 (MP), CONICYT PhD Program Fellowship (FC), Spanish
Ministerio de Educaci\'on y Ciencia (Project MTM2005-09183) and
Junta de Castilla y Le\'on (Excellence Project VA013C05) (LMN). LMN
also thanks the Mecesup Project USA0108 for making possible his
visit to the University of Santiago de Chile, and Department of
Physics of this University for hospitality.


\end{document}